# Decision Support Tools for Cloud Migration in the Enterprise

Ali Khajeh-Hosseini[1], Ian Sommerville[1], Jurgen Bogaerts[2], Pradeep Teregowda[3]

[1]Cloud Computing Co-laboratory  
School of Computer Science  
University of St Andrews, UK  
{akh, ifs}@cs.st-andrews.ac.uk

[2]Economics & Management Faculty  
HUBrussel  
Brussels, Belgium  
jurgen.bogaerts@student.hubrussel.be

[3]Computer Science & Engineering  
Pennsylvania State University  
Pennsylvania, USA  
pbt105@psu.edu

*Abstract* — **This paper describes two tools that aim to support decision making during the migration of IT systems to the cloud. The first is a modeling tool that produces cost estimates of using public IaaS clouds. The tool enables IT architects to model their applications, data and infrastructure requirements in addition to their computational resource usage patterns. The tool can be used to compare the cost of different cloud providers, deployment options and usage scenarios. The second tool is a spreadsheet that outlines the benefits and risks of using IaaS clouds from an enterprise perspective; this tool provides a starting point for risk assessment. Two case studies were used to evaluate the tools. The tools were useful as they informed decision makers about the costs, benefits and risks of using the cloud.**

*Cloud migration; cloud adoption; IaaS; enterprise cloud computing; cost modeling; risk assessment*

## I. INTRODUCTION

Enterprises are interested in using public clouds, especially infrastructure-as-a-service (IaaS), because of its scalability, flexibility and apparent cheapness. However, unlike start-ups that develop systems from scratch, enterprises often have to deal with so-called 'brownfield development' where new systems have to inter-operate with a range of existing systems [1]. Decisions to migrate existing systems to public IaaS clouds can be complicated as evaluating the benefits, risks and costs of using cloud computing is far from straightforward [2]. Organizational and socio-technical factors must also be considered during the decision making process as the shift towards the cloud is likely to result in noticeable changes to how systems are developed and supported [3].

There is a need for guidelines and decision support tools for enterprises that are considering migrating their IT systems to the cloud (*cloud migration*). Cloud providers are attempting to address this demand with whitepapers offering advice (e.g. [4] and [5]), while IT consultancies are offering frameworks (e.g. [6]) and assessment tools (e.g. [7] and [8]) to support decision makers. However, such tools are either marketing tools or they are not widely available as they are based on closed proprietary technologies that are often accompanied by expensive consultancy contracts. Therefore, the original contribution of this paper is to describe and evaluate two impartial tools that aim to support cloud migration decisions.

The first tool is a modeling tool that produces cost estimates of using public IaaS clouds. This enables IT architects to model their applications, data and infrastructure requirements in addition to their computational resource usage patterns. It can be used to compare the cost of different cloud providers, deployment options and usage scenarios. The second tool is a spreadsheet that outlines the benefits and risks of using IaaS clouds from an enterprise perspective, and provides a starting point for risk assessment.

The tools were evaluated using two case studies representing different types of systems. The first case study involved an academic digital library and search engine (called *CiteSeer$^x$*) that indexes a continually growing set of documents from the web. This system currently contains over 1.5 million documents and has around 2 million hits per day. This represents a highly technical and automated system that is managed by a small team, which can be likened to a small enterprise that is free from the organizational hierarchy and overheads of large enterprises. The second case study involved the IT systems of the European R&D division of a large media corporation that has over 20,000 employees worldwide. This case represents a typical enterprise division with its own independently-managed systems that are part of a large inter-connected corporate IT environment. The tools were useful in both case studies as they informed decision makers about the costs, benefits and risks of using the cloud.

The remainder of this paper is structured as follows: Sections 2 and 3 describe the tools; Section 4 provides the case study results and discusses their significance; Section 5 reviews related work; while Section 6 concludes the paper and describes our future work.

## II. COST MODELING

Enterprises have to consider several types of costs during cloud migration, including IT infrastructure, data centre equipment and real estate, software licenses, systems engineering and software changes, staff costs etc. Most of these costs can be calculated with spreadsheets, and indeed cloud providers such as Amazon have created spreadsheets for cost comparisons. However, the cloud's utility billing model has a certain degree of uncertainty that makes it complicated to use spreadsheets to calculate the infrastructure costs of using IaaS clouds. This uncertainty relates to: 1) the actual resources consumed by a system, which are determined by its load; 2) the deployment options used by a system, which can affect its costs as things like data transfer are more expensive between clouds compared to data transfer within clouds; and 3) cloud providers' prices, which can change at short notice.

Our cost modeling tool simplifies the process of obtaining infrastructure cost estimates for the deployment of an existing/planned IT system across different clouds. The tool can be used by IT architects or systems engineers to create a model of their system. The tool processes the model, taking into account prices from cloud providers, and outputs a report showing how the costs would vary over time.

The cost modeling tool enables users to model the infrastructure of their IT system by extending UML deployment diagrams using a custom UML profile. UML deployment diagrams have a fairly simple notation, and enable users to model the deployment of software artifacts onto hardware nodes [9]. The tool's 'cloud deployment' UML profile was installed as part of the Eclipse IDE; it extends UML deployment diagrams to include:

- *Virtual Machine*: has an operating system, and either a server type (e.g. AWS.OnDemand.Standard.Small) or server specifications (e.g. CPU clock rate and RAM).
- *Virtual Storage*: represents persistent storage and can have a type (e.g. AWS.EBS or AWS.S3) in addition to a size (e.g. 100GB) and the number of input and output requests that are expected per month.
- *Application*: represents software applications that are deployed on virtual machines.
- *Data*: represents application data that is deployed on virtual storage.
- *Database*: represents hosted databases such as Amazon's Relational Database Service or Microsoft's SQL Azure.
- *Remote Node*: represents nodes outside of the cloud such as in-house servers or desktop PCs.
- *Communication Path*: represents data transfer between any pair of nodes.
- *Deployment*: represents the deployment of applications onto virtual machines, or data onto virtual storage.

Once a basic model has been created with Eclipse's drag & drop GUI, users specify which cloud provider they would like to use for each node, and how much computational resources they need. The following computational resources are supported: running hours for virtual machines; storage; number of data I/O requests for storage; and the amount of data in and out of nodes. There can be other costs associated with running a system in the cloud, e.g. the cost of a static IP address; however, such costs are usually insignificant.

One of the key benefits of using the cloud is elasticity and we have developed a simple notation to allow elasticity requirements to be expressed. The tool enables users to define a baseline usage for each resource. Variations to this baseline can be defined using 'elasticity patterns' that are expressed in natural language. Each pattern can either be temporary or permanent. A temporary pattern is only applied during the month(s) that it is applicable, and can be used to define temporary peaks or drops in usage. In contrast, the resource usage that is changed by a permanent pattern is persistent. Therefore, permanent patterns can be used to define patterns of linear or exponential resource increases or decreases. A pattern is defined as follows:

*[temp/perm]: every [months] on [days] [variation][number]*
Where *months*, *days*, *variation* and *number* can be:

| Months | Days | Variation | Number |
|---|---|---|---|
| month | [empty] | + | float or integer |
| jan-dec | everyday | - | |
| | weekdays | * | |
| | weekends | / | |
| | 01-31 | ^ | |
| | mon-sun | | |

For example, the following patterns describe a scenario where initially 100GB of storage is required; every month this is increased by 10GB; during weekends between June and August, the required storage is halved; and every December between 25th and the 30th, it is doubled.

```
Baseline: 100, Patterns:
perm: every month +10,
temp: every jun-aug on weekends /2,
temp: every dec on 25-30 * 2
```

After a cloud deployment model has been created and elasticity patterns have been defined, users set a start and end date for the cost simulation to be performed. The tool then starts the simulation, representing the model as a directed cyclic graph. The usage patterns of each node and edge in the graph are processed for each month between the simulation start and end dates. The total resource usage of each node is then multiplied by the per-unit cost of that resource, depending on which cloud provider is specified by the user. The per-unit price is retrieved from an XML file that stores cloud providers' prices. It currently contains over 600 prices from AWS, MS Azure, FlexiScale, Rackspace, GoGrid, and ReliaCloud. Other providers and prices can easily be added.

Finally, the tool generates a report showing how the cost of the system would change over time. The report is a webpage with embedded graphs, tables and a zoomable version of the model, which becomes useful when dealing with large system models. The tool can also export the full costing details as a CSV table for further analysis in Excel.

Each model can be divided into different groups, and the report provides a detailed cost breakdown for each group. A group can represent a department, an organization or an entire/part of a system. This enables architects to evaluate different deployment options for a system and see which is the cheapest. For example, system architects can investigate the costs of duplicating parts of the system on a different cloud for increased availability.

### III. BENEFITS AND RISKS ASSESSMENT

The potential cost savings of using cloud computing have to be examined in the wider context of other benefits and risks. It can be difficult or meaningless to quantify indirect cost savings of, say, the improved time-to-market or flexibility provided by using public IaaS clouds. From an enterprise perspective, costs are important but so too are customer relationships, public image, flexibility, business continuity and compliance. Therefore, the motivation for creating a benefits and risks assessment tool is to inform decision makers by bringing these issues together.

This paper uses the following definition of a benefit: an advantage to the enterprise over its status quo provided by using public IaaS clouds. The ISO/IEC Guide 73:2002 defines risk as the "combination of the probability of an event and its consequence". In this paper, risks are seen as events that might occur if public IaaS clouds are used.

The benefits and risks of using public IaaS clouds were identified by reviewing over 50 academic papers and industry reports. Examples are shown in Tables 1 and 2. The benefits/risks were categorized as organizational, legal, security, technical or financial. Due to space limitations, Table 1 has 10 out of the 19 identified benefits, and Table 2 has 20 out of the 39 identified risks (full tables are available from ShopForCloud.com as a Google Docs spreadsheet).

The spreadsheet provides a starting point for risk assessment as it identifies the risks and describes their potential consequences, some mitigation approaches and potential indicators. Users can go through the spreadsheet and rate each item as unimportant, little important, moderately important, important, or very important from their perspective. This type of scaling is called the *Likert scale* and is often used in survey-based research.

The spreadsheet includes benefits and risks that were initially identified in the IT outsourcing literature but could equally apply to cloud migration. As part of a comprehensive analysis and survey of the IT outsourcing literature, Dibbern et al. [10] point out that initially, researchers in IT outsourcing were concerned with *why* organizations outsourced their IT (e.g. major factors involved in the decisions, advantages and disadvantages of outsourcing). Later on, researchers focused on *how* organizations outsourced their IT (e.g. evaluating different vendors and structuring IT outsourcing contracts).

The IT outsourcing literature has shown that IT outsourcing does not always meet expectations. Lacity and Hirschheim [11] analyzed the extent to which the expectations of fourteen Fortune 500 companies engaged in IT outsourcing were met. The study revealed that many of the outsourcing success stories portrayed in the literature painted an inaccurate picture. For example, outsourcing contracts that promised clients cost savings of 10 to 50% of their IS costs over the life-time of the contract were often anticipated savings that were not actually achieved. Lacity and Hirschheim [11] found that in some cases, in-house IT departments could make similar cost savings through standardization, consolidation and internal charge-back mechanisms, but were prevented from implementing such strategies by internal politics and organizational culture.

Dibbern et al. [10] wonder if IT outsourcing is "nothing more than a pendulum" that started with organizations creating internal IS departments only to realize that outsourcing can be more beneficial, but after going through several outsourcing contracts, discovered that it is unsatisfactory. Therefore, they are bringing their outsourced IT systems back in-house due to poor service levels, changes in strategic direction and failed cost saving promises [12].

We see cloud computing as another swing in the IT outsourcing pendulum, where organizations are seeking to outsource their IT infrastructure due to the potentially cost effective, scalable and reliable services provided by cloud providers. However, there is a key difference between cloud computing and IT outsourcing: the cloud's self-service model and the lack of fixed long-term contracts give more control and flexibility to clients compared to traditional IT outsourcing. For cloud computing not to result in another swing-back of the IT outsourcing pendulum, it is important that organizations consider the risks mentioned in Table 2 during their decision making process.

TABLE I. EXAMPLES OF THE BENEFITS OF USING PUBLIC IAAS CLOUDS, COMPLETE TABLE AVAILABLE FROM SHOPFORCLOUD.COM

| ID | Benefit | References |
|---|---|---|
| B1 | **Technical:** Fast access to additional computational resources and specialized skills (e.g. IT specialists who build and maintain clouds). Results in quicker system deployment times. | [13], [10], [14], [15] |
| B2* | **Technical:** Ability to address volatile demand patterns and the flexibility to scale-up/down resource usage without discontinuity or service interruption. Reduced risk of over/under provisioning infrastructure resources. | [16], [17], [10], [18] |
| B3 | **Technical:** Reduced run/response time due to the ability to acquire vast computational resources for short time periods, e.g. a batch job taking 1000hrs can be done in 1hr using 1000 servers for the same cost. This can lead to a reduced time to market. | [18], [16], [19], [20] |
| B4* | **Technical:** Anywhere/anytime/any device (laptop, mobile etc.) access to computational resources & applications can be setup without too much effort. This simplifies collaboration amongst users & simplifies application support and maintenance. | [21], [5] |
| B7 | **Technical:** Simplified and cheaper provisioning of disaster recovery and business continuity plans due to geo-distribution and replication facilities provided by cloud providers. | [19], [22] |
| B9 | **Financial:** Reduced costs due to more efficient operations and less infrastructure maintenance costs but also due to economies of scale that can be achieved by cloud providers. | [23], [13], [10] |
| B11 | **Financial:** Reduced need for capital investment and the ability to transform fixed costs into variable costs. This might simplify cash-flow management. | [19], [18], [3] |
| B14 | **Organizational:** Ability to focus on core business activities and free-up management and IT personnel from mundane tasks (such as hardware maintenance) so that they can focus on value-added activities. | [17], [13], [10], [15] |
| B16* | **Organizational:** Opportunity to offer new products or services or trial products to gauge the level of interest from customers. | [3] |
| B18* | **Organizational:** Devolution of decision making on IT requirements to operational units. Variable provision in different parts of the organization (this could also be a risk) | [3] |

\* Can also apply to private clouds

TABLE II. EXAMPLES OF THE RISKS OF USING PUBLIC IAAS CLOUDS, COMPLETE TABLE AVAILABLE FROM SHOPFORCLOUD.COM

| ID | Risk | Mitigation approaches & potential indicators | References |
|---|---|---|---|
| R1 | **Organizational:** Loss of governance and control over resources (both physical control and managerial), might lead to unclear roles and responsibilities, e.g. users can purchase computing resources using their credit cards without explicit approval from central IT. | Clarify roles and responsibilities before cloud adoption. | [3], [24], [25], [10], [14] |

| ID | Risk | Mitigation approaches & potential indicators | References |
|---|---|---|---|
| R3* | **Organizational:** Reduced staff productivity during the migration as changes to staff work (e.g. staff getting less satisfying work) and job uncertainty (or departmental downsizing) leads to low staff morale and anxiety spreading in the organization. | Ensure that experts are not dismissed and involve them in the migration project so that they get a sense of ownership. Indicator: rumors spreading in the organization about future job uncertainty. | [15], [3] |
| R5 | **Organizational:** Managing a system deployed on several clouds can take extra management effort compared to deploying systems in-house (e.g. to manage the relationship with the cloud providers, deal with problems, deal with changes to cloud services). This is one of the hidden costs of deploying systems on the cloud. | Make management aware of the extra effort that might be required. Indicator: using several cloud providers for a system, cloud providers having different types of support mechanisms. | [10], [26], [27] |
| R7 | **Organizational:** Changes to cloud providers' services or acquisitions by another company that changes/terminates services. | Use multiple providers. Indicator: using a small provider that might be acquired by bigger companies. | [24] |
| R11* | **Organizational:** Resistance to change resulting from organizational politics and changes to people's work. | Use insights from organizational change management and involve key stakeholders in the adoption process. Indicator: organizational gossip. | [17], [3], [25] |
| R39 | **Organizational:** Mismatch between existing incident handling procedures and cloud providers' procedures. Lack of information or no access to a cloud's vulnerability information or incident report data. Leads to limited responses from an organization in case of incidents. | Check cloud provider's SLA and ensure that it has well-defined incident classification schemes and reporting procedures (e.g. what is reported, how fast it is reported, to whom it is reported). Indicator: lack of incident handling information in the SLA. | [28] |
| R12* | **Legal:** Unusable software licenses on the cloud due to the license using traditional per-seat or per-CPU licensing agreements etc. | Check all software license agreements. Indicator: using software that requires physical software locks. | [24], [29] |
| R13 | **Legal:** Non-compliance with regulations that require informed consent from users when dealing with personal data (e.g. the first principle in the UK's Data Protection Act 1998). | Avoid risk by explicitly asking users to give consent. Refer to [30] for a detailed review of standard contracts that are often used by cloud providers. | [31] |
| R15 | **Legal:** Lack of information on jurisdictions used for data storage and processing. Leads to non-compliance with regulations that require certain types of data to be kept in national boundaries (e.g. the eighth principle in the UK's Data Protection Act 1998 that requires personal data to be kept within the EEA). | Avoid risk by using data centers within the required jurisdiction. Indicator: cloud providers that do not disclose the country of their data centers. | [31] |
| R16 | **Legal:** Non-compliance with data confidentiality regulations. Unauthorized access to data by cloud providers. | Use encrypted data transfer and storage. When using IaaS, follow AWS's security guidelines. Indicator: cloud providers not supporting encrypted data transfer. | [31], [16], [32], [33] |
| R18 | **Legal:** Non-compliance with industry regulations, such as the Financial Services Authority regulations in UK, and the following regulations in US: Health Insurance Portability and Accountability Act, Federal Information Security Management Act, Payment Card Industry Data Security Standards (PCI DSS), Sarbanes-Oxley (SOX) and Statement on Auditing Standards No 70 (SAS 70). | Check compliance with the auditors and cloud providers. When using IaaS and dealing with data that is protected by HIPAA, follow AWS's guidelines. Indicator: lack of regulated customers using the cloud. | [31], [24] |
| R21* | **Security:** Denial of service attacks. Leads to unavailability of resources and increases cloud usage bills. | Use network monitoring tools (although some providers do not allow this) or monitor applications from outside of the cloud. Indicator: slower performance than expected, reports from customers that services are unavailable. | [24], [34], [32], [35] |
| R23 | **Security:** Interception of infrastructure management (API) messages and data in transit. This could lead to the infrastructure being manipulated by third parties. | Use secure communication protocols and multi-factor authentication. When using IaaS, follow AWS's security guidelines. | [24] |
| R25* | **Security:** Browser vulnerabilities become more significant. | Ensure browser updates are deployed in a timely manner. | [34], [36] |
| R26* | **Technical:** Major service interruption resulting in extensive outages and unavailability of services or loss of data. | Use multiple cloud providers, monitor applications from outside the cloud. Replicating the system across multiple clouds has associated costs and technical challenges. | [16], [37], [3], [32], [21] |
| R27 | **Technical:** Data lock-in for SaaS/PaaS and system lock-in for IaaS. | For IaaS lock-in, use middleware that is compatible with multiple clouds (e.g. RightScale). See [24] (p.26) for other mitigation approaches. Indicator: lack of interest from providers to participate in standardization efforts. | [16], [14], [24] |
| R28 | **Technical:** Performance is worse than expected (e.g. CPU clock rate, I/O and network data-transfer and latency rates). It might be difficult to prove to the cloud provider that their system performance is not as good as they promised in their SLA as the workload of the servers and the network can be highly variable in a cloud. This might lead to disputes and litigation. | Use benchmarking tools to investigate performance of the cloud under investigation for decision making. Rent more VMs or higher spec ones to deal with slow CPU clock rates, use physical disk shipping to reduce effects of network latency/transfer rates. Use third party monitoring tools to independently verify the system performance. | [16], [38], [39], [26], [14] |
| R31 | **Technical:** Interoperability issues between clouds as there are incompatibilities between cloud providers' platforms. | Use cloud middleware (e.g. RightScale) to ease interoperability issues. | [24] |
| R34 | **Financial:** Actual costs may be different from estimates, this can be caused by inaccurate resource estimates, providers changing their prices, or inferior performance (e.g. due to over-utilized servers) resulting in the need for more resources than expected. | Monitor existing resource usage and use estimation tools to obtain accurate cost estimates of deploying IT systems on the cloud. Check results of performance benchmarks. | [2], [32], [26] |
| R36* | **Financial:** Increased costs due to complex integrations. Inability to reduce costs due to unrealizable reductions in sys-support staff. | Investigate system integration issues upfront, avoid migrating highly interconnected systems initially. | [13], [5], [40] |

\* Can also apply to private clouds

## IV. CASE STUDIES

### A. Digital Library and Search Engine

This case study represents a highly technical and automated system that is managed by a small team. The team can be likened to a small enterprise that is free from the organizational hierarchy and overheads of large enterprises. The system under investigation in this case study is a digital library and search engine (called *CiteSeer$^x$*) that indexes academic papers, and enables users to search and access papers via the web. This system uses a service-oriented architecture and has the following components: a web application that is the main interface between the system and its users; document crawling and metadata extraction; ingestion, which processes the crawled documents and updates the system repository; a maintenance service that updates the indexes and generates relevant statistics; and finally a data backup and replication service. These components are deployed on 15 servers that are housed in a university data center. The system has around 2 million hits per day and contains over 1.5 million documents requiring around 2TB of storage.

#### 1) Cost Modeling

The applications and infrastructure of the system were modeled using the cost modeling tool. Based on historic data, a number of elasticity patterns were also created to model the growing resource needs of the system. These patterns included a 15GB/month increase in data out of the system (caused by increasing number of visitors), and a 17GB/month increase in the size of the document repository.

The cost of using different cloud providers was investigated, and as shown in Table 3, Amazon Web Services (AWS) was found to be the cheapest. The significant price difference between the providers was unexpected as it is argued that cloud providers compete on price [38].

TABLE III.  COST OF DIFFERENT CLOUD PROVIDERS

| Cost ($) | AWS US-East | FlexiScale | Rackspace |
|---|---|---|---|
| 1st month | 18,980 | 5,060 | 6,550 |
| Monthly avg. | 1,916 | 5,151 | 6,732 |
| Total, 3 years | 85,950 | 185,345 | 242,170 |
| Difference with AWS | | +2x | +3x |

Cost modeling highlighted some interesting points:
1. Performance must also be considered but the performance vs. cost trade-off of IaaS clouds is an open research challenge as traditional benchmarking approaches are unlikely to be adequate since their results depend on the applications running and network-load of the cloud at that time; hence they are not easily generalizable.
2. The cost breakdown of AWS showed that 66% was for VMs, 20% was for data transferred out of the system, 10% was for storage, and 4% was for storage I/O requests. This information can be used to optimize the system for cost, e.g. switching off VMs when they are not in use is a simple but effective cost cutting technique.
3. Using AWS's S3 storage would cost around $4,000 more than using EBS (over the 3 years). This is due to the increased cost of storage I/O requests from S3. However, this estimate might be inaccurate as it is difficult to calculate the storage I/O requests for S3 by looking at disk I/O figures from Linux's `iostat` command.
4. Storage I/O is priced differently by providers: AWS charge for number of I/O requests to storage, FlexiScale charge for the amount of data transferred to/from storage, Rackspace do not charge for input requests for files over 250KB in size, and GoGrid do not charge at all.

#### 2) Benefits and Risks Assessment

The benefits/risks spreadsheet was used by the technical director of the digital library and search engine. The exercise took around one hour; 7 benefits and 13 risks were identified as important. The results are discussed in Section 4.C.

### B. R&D Division of a Media Corporation

This case study represents a typical enterprise division with its own independently-managed systems that are part of a large inter-connected corporate IT environment. The systems under investigation in this case study belong to the European R&D division of a large media corporation that has over 20,000 employees worldwide. This division is responsible for research and development in software applications that are used by the corporation. The division has around 40 office and management personnel, each with a laptop that is used for office applications and occasional consultation of software development functionality. There are also around 120 engineers, each with a workstation that is mainly used for software development and testing.

The R&D division uses a range of applications including IBM ClearQuest (software change management), IBM DOORS (requirements management), IBM ClearCase (software configuration management), Klocwork (static source code analysis) as well as a number of databases (SQL Server) and custom-made websites. These applications are deployed on 9 heavy-duty servers and 2 network storage systems that are housed in the local office.

#### 1) Cost Modeling

The R&D division is currently thinking about its future infrastructure strategy and is interested in finding out the infrastructure costs of using AWS so that it can compare it with other options (e.g Eucalyptus private cloud). The current setup of their applications and infrastructure was modeled using the cost modeling tool. The current setup is based on a client-server architecture, where users have powerful machines that are used for most of their work including code compilation. This setup results in the servers having a moderate load. The workstations are not usually under a heavy load but at times, users have to enlist the workstation of their neighbors to speedup code compilation.

The division is considering moving to a thin client setup where each engineer would have a standard office laptop from which they could connect to centralized servers from their home, office or a customer site. This new setup would include the current servers as well as new session servers for users to login to and work, and compute servers that are used to off-load heavy processing or host applications that have dedicated resource needs. The costs of three options were investigated: 1) using AWS instances in a non-elastic manner by leaving them on 24x7; 2) using AWS instances in an elastic manner where on working days, the session servers would be on during day-time and compute servers would be on during nigh-time; and 3) using AWS instances in an elastic manner as in option 2 but giving each user a small instance rather than using a few large instances to host all user sessions. Using option 3 would mean that a user could switch on/off their instance at anytime without worrying about who else is using that instance, hence the small instances would be on for 8hours per working day. The costs of the different deployment options are shown in Table 5.

TABLE IV. COST OF DIFFERENT DEPLOYMENT OPTIONS ON AWS-EU

| Cost ($) | Non-elastic | Elastic | Elastic, small instances |
|---|---|---|---|
| 1st month | 67,350 | 65,430 | 75,260 |
| Monthly average | 6,259 | 4,344 | 4,175 |
| Total for 3 years | 286,415 | 217,470 | 221,385 |

Cost modeling highlighted some interesting points:
1. Typical enterprise divisions do not currently need to monitor their actual computational resource consumption in the level of detail that is required for cost modeling. Since servers are either procured upfront or leased from an IT vendor under a provisioning and support contract, the actual resource consumption does not affect costs in a significant manner. This issue highlights the need for monitoring tools that provide resource usage estimates that can be fed into cost modeling tools.
2. In contrast to popular belief, using AWS can have a fairly high start-up cost, which is used to reserve instances that reduce monthly costs in the long-term. In this case, it was around 30% of the total costs over the 3 years.
3. The elasticity of the cloud can be used to reduce costs. In this case, the difference between the non-elastic and elastic setup would be around $70,000 over the 3 years.
4. Different deployment options have to explored to find the cheapest. In this case, using 4 high-memory quadruple-extra-large instances costs around $4,000 less than using 120 small instances (one for every engineer). However, using small instances is more flexible as instances can be switched on when individual engineers need them, whereas a large instance has to be switched on even if only one engineer is using it. Whenever possible, enterprises should use small instances that enable them to increase or decrease their resource consumption by small chunks to avoid having under-utilized servers.

*2) Benefits and Risks Assessment*

The benefits and risks spreadsheet was used by the IT manager of the R&D division who provided a local view of the benefits and risks. The exercise took around two and a half hours and included discussions with one of their senior software engineers. The spreadsheet was also used by one of the IT managers who works directly with corporation's CIO; this provided a corporate view of the benefits and risks. The technical analysis revealed 8 important benefits and 25 important risks; the organizational analysis suggested 4 important benefits and 15 important risks.

*C. Discussion*

There are many benefits and risks involved in using public IaaS clouds. To get a holistic picture of the benefits and risk from an enterprise perspective, the weighted average of the benefits/risks can be calculated and charted on a radar graph, as shown in Figures 1 and 2. The weighted average can be calculated by multiplying the number of benefits/risks in each category (organizational, legal, security, technical or financial) by the weight of each benefit/risk (unimportant = 1 … very important = 5), and dividing the result by the total number of benefits/risks in that category.

Figure 1 shows the weighted average of the benefits for the case studies. It shows that in the case of the digital library, the technical benefits of using public IaaS clouds were more important than the organizational and financial benefits. Hence, the technical ability to deal with volatile demand patterns and cater for a growing number of users would be one of the main motivations for using the cloud. In the case of the media corporation, the R&D division also views the technical benefits as more important than the organizational and financial ones. From their perspective, the simplified provisioning of computational resources, and anywhere/anytime access to resources are some of the motivations for using the cloud. Whereas it is clear that their corporate IT department views financial and organizational benefits as more important than technical ones.

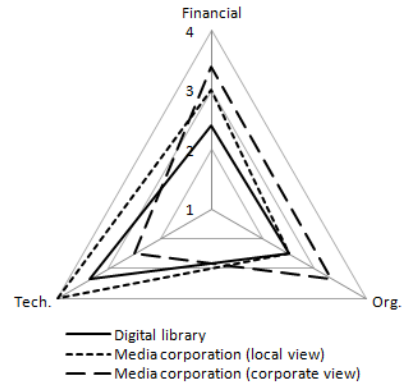

Figure 1: The importance of the different types of benefits of cloud migration in case studies presented in this paper (1=unimportant, 2=little important, 3=moderately important, 4=important, 5=very important)

Figure 2 shows the weighted average of the risks of cloud migration. In the digital library case study, it is clear that the main risks are financial, technical and security related; legal and organizational risks are not important as are not relevant to small enterprises that deal with non-sensitive data. The R&D division of the media corporation has a good understanding of their local systems and appreciates the importance of the different types of risks, whereas their corporate IT department is mostly concerned with the organizational and legal risks. This is understandable as corporate IT are probably best equipped to deal with those risks, however, they are not too concerned with financial risks as budgets are probably managed by local divisions.

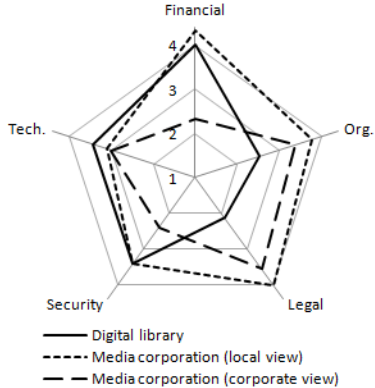

Figure 2: The importance of the different types of risks of cloud migration

The case studies show that there are multiple perspectives on the benefits/risks of cloud migration, even within a single enterprise, different divisions have different views. Motivations and concerns of different stakeholders have to be considered during cloud migration decisions and in the future we are going to develop a lightweight method that enables these issues to be further explored and mapped-out. It is hoped that this method will support decision makers by enabling them to visualize and understand the benefits/risks of cloud migration from different perspectives.

## V. RELATED WORK

Little academic work has been done to investigate the factors involved in cloud migration. Therefore, we have so far taken a multiple case study approach to understand the factors and evaluate our decision support tools. Our first case study investigated the feasibility of cloud migration in an SME in the oil & gas industry [3]. This case study was exploratory in nature, and provided insights about socio-technical factors involved in cloud migration. Based upon those insights, we described some of the challenges that enterprises face during cloud migration, and introduced the Cloud Adoption Toolkit that aims to support decision makers in [2]. The two tools described here are part of the toolkit.

Our second case study, also described in [2], evaluated the cost modeling tool by using it to compare the costs of buying physical servers with leasing resources from Amazon Web Services. Furthermore, this case study investigated the effects on infrastructure costs if a cloud provider increases or decreases its prices sometime in the future. The cost modeling tool was initially described in [2] but was also briefly described in this paper as it was used in the case studies presented here. However, the tool is used differently in this paper compared to [2]; namely, it is used to compare the costs of different cloud providers, and different deployment options (e.g. different types of instances).

Others have also investigated the costs of using the cloud via individual case studies (e.g. [41], [42], and [43]). Some work has also been done on the cost modeling aspects of cloud computing from a user perspective (e.g. [44] and [45]), and the performance vs. cost trade-off of using different clouds [46]. However, we argue that the cost modeling tool presented in this paper is an improvement on those works as it can be used in practice to model: IT infrastructure, deployment options, elasticity patterns, different cloud providers' cost models, and changes to their cost models.

In terms of the benefits and risks of using cloud computing, the references in Section 3 point to related work, which has until now been sparsely reported in various academic papers and industry reports. The comprehensive review presented in this paper builds upon those existing works, and presents the benefits and risks in a tool that provides a starting point for risk assessment. It should also be noted that previous works such as [47] and [48] have mostly focused on the security risks of using cloud computing, whereas this paper argues that other risks must also be considered during the decision making process. This paper has shown that the IT outsourcing literature, which has until now been overlooked in the cloud computing literature, can provide a rich source of information about the organizational benefits and risks of using the cloud.

## VI. CONCLUSION

This paper described two tools (cost modeling, and benefits and risks assessment) that aim to support decision making during the migration of IT systems to the public IaaS clouds. The cost modeling tool enables IT architects to model their applications, data and infrastructure requirements in addition to their computational resource usage patterns. The benefits and risks spreadsheet provides a starting point for risk assessment as it outlines the organizational, legal, security, technical and financial benefits and risks of using IaaS clouds from an enterprise perspective. The tools were evaluated using two case studies representing a technical system managed by a small team, and a corporate enterprise system. The first case represented a small enterprise that is free from the organizational hierarchy and overheads of large enterprises. The second case study represented a typical enterprise division that has its own independently-managed systems, which are part of a large inter-connected corporate IT environment.

One of the limitations of this paper is that it only discusses infrastructure costs of using public IaaS clouds. Cost modeling has to be used in conjunction with project management and software cost estimation techniques to enable the full costs of cloud migration to be investigated. However, this was not possible in this paper as the enterprises involved in the case studies have not made their cloud migration decision yet. The digital library might use public IaaS clouds for a subset of their system components. The media corporation's R&D division is unlikely to migrate its systems to public IaaS clouds due to the identified risks; however, it might use the cloud for specific functions such as offsite backup and disaster recovery. It is unlikely that enterprises will migrate existing IT systems to the cloud before the performance of their existing infrastructure becomes unacceptable. However, new systems or additions to existing systems might be good candidates for the cloud as they are unlikely to suffer from migration issues.

Despite the previously mentioned shortcomings of this paper, the enterprises involved in the case studies found the tools useful as they informed decision makers about the costs, benefits and risks of using public IaaS clouds. We plan to perform a longitudinal study of the enterprises involved in the case studies, and as part of our future work, report on their cloud migration decisions and its effects on their systems and organizations. The tools described in this paper and other tools that we are currently developing, will be made available from www.ShopForCloud.com.

ACKNOWLEDGMENT

We thank the SICSA and the EPSRC (grant numbers EP/H042644/1 and EP/F001096/1) for funding this work. We also thank our colleagues, especially David Greenwood, at the UK's LSCITS initiative for their comments.